# *The non-capacitor model of leaky integrate-and-fire $VO_2$ neuron with the thermal mechanism of the membrane potential*

**A A Velichko[1], M A Belyaev[1], D V Ryabokon[2,3] and S D Khanin[2,3]**

[1]Institute of Physics and Technology, Petrozavodsk State University, 33 Lenin str., 185910, Petrozavodsk, Russia
[2]Marshal of the Soviet Union Budennyi Military Academy of Communications, 3 Tikhoretsky pr., 194064 St. Petersburg, Russia
[3]Gertsen Russian State Pedagogical University, 48 nab. Moika river, 191186, St. Petersburg, Russia

E-mail: velichko@petrsu.ru

***Abstract.*** The study presents a numerical model of leaky integrate-and-fire neuron created on the basis of $VO_2$ switch. The analogue of the membrane potential in the model is the temperature of the switch channel, and the action potential from neighbouring neurons propagates along the substrate in the form of thermal pulses. We simulated the operation of three neurons and demonstrated that the total effect happens due to interference of thermal waves in the region of the neuron switching channel. The thermal mechanism of the threshold function operates due to the effect of electrical switching, and the magnitude (temperature) of the threshold can vary by external voltage. The neuron circuit does not contain capacitor, making it possible to produce a network with a high density of components, and has the potential for 3D integration due to the thermal mechanism of neurons interaction.

## *1. Introduction*

In recent decades, fundamental and applied research in the field of artificial intelligence has been actively conducted, where artificial neural networks (ANNs) play the leading role [1,2]. The main efforts are focused at developing new architectures, optimal learning algorithms and ways to improve the accuracy of ANN operation [3,4]. One of the promising research directions is the development of biosimilar neurons imitating the spikes of action potential in spike neural networks (SNN) or in the third generation ANN [5,6]. Several models of spike neurons can be distinguished in the literature: leaky integrate-and-fire (LIF), FitzHugh-Nagumo, Hindmarsh-Rose, Morris-Lecar, Hodgkin-Huxley, Izhikevich [7]. The size of a neuron is one of the most important parameters of any neuromorphic microcircuit, as it determines the number of neurons that can be integrated in a single die. A larger number of neurons would allow the solution of more complex cognitive tasks. Currently, there are microcircuits containing $10^6$ neurons, while the human brain contains about $10^{11}$ neurons [8]. Therefore, reducing the size of a neuron and manufacturing more compact neural networks with the possibility of 3D integration are the relevant tasks, especially for integrate-and-fire neurons, where a capacitor is used as an integrating element to add the actions from all inputs of the neuron. One of the methods of miniaturization is the reduction of elements in the circuit, and in particular, capacitors that occupy a significant area. It can be implemented by employing new physical effects in the mechanism of neuron functioning. One of the promising materials for neuromorphic devices is vanadium dioxide. Having a metal insulator transition and the effect of electrical switching, this material is a good alternative for the creation of neuromorphic, logical, and oscillatory devices [9–11].

In the current study, we investigate one of the simplest biosimilar models of a neuron – LIF neuron, based on a $VO_2$ switch. The temperature of the switch channel is used as an analogue of the membrane potential. The action potential from neighbouring neurons propagates along the substrate in the form of thermal pulses. The summation of neurons interaction and the effect on the membrane potential happens in the region of the neuron switching channel due to interference of thermal waves. The thermal mechanism of the threshold function is caused by the effect of electrical switching.

The simplest circuit of a neuron is one $VO_2$ switch and one load resistor connected to a power source, without a capacitor. The study presents the results of numerical simulation of a circuit of three interacting $VO_2$ neurons and explores the concept of the architecture of a pulsed neural network with a thermal mechanism of the membrane potential and threshold function.

## *2. Circuit of the $VO_2$ neuron*

The neuron circuit is presented in Figure 1(a). The main element of a neuron is a two-electrode switch based on vanadium dioxide. This material has a metal-insulator phase transition with a sharp change in resistivity, up to 5 orders of magnitude, at a phase transition temperature $T_{th}$ ~ 340 K, which determines the presence of the electric switching effect in $VO_2$ structures [12]. The simulation of two-electrode planar switching structures of Au-$VO_2$-Au was performed in the COMSOL Multiphysics modelling software, with the physical parameters of the $VO_2$ film presented in [13], the electrode capacitance was not taken into account. The planar dimensions of the gold contacts and the $VO_2$ film were 1 μm x 1 μm, and the thickness was 100 nm. Sapphire was used as the substrate material (substrate size 20 μm x 20 μm x 20 μm). When simulating the *I–V* characteristic of the switch, a linearly increasing voltage was applied to a separate $V_{sw}$ switch (voltage slew rate ~ $10^3$ V / s) in the range from 0 to 6 V, then the voltage was decreased to 0 V at the same rate. *I–V* characteristic of the switch, dependence of the current of the $I_{sw}$ switch on the voltage $V_{sw}$ at ambient temperature $T_0$ = 300 K is presented in Figure 1(b).

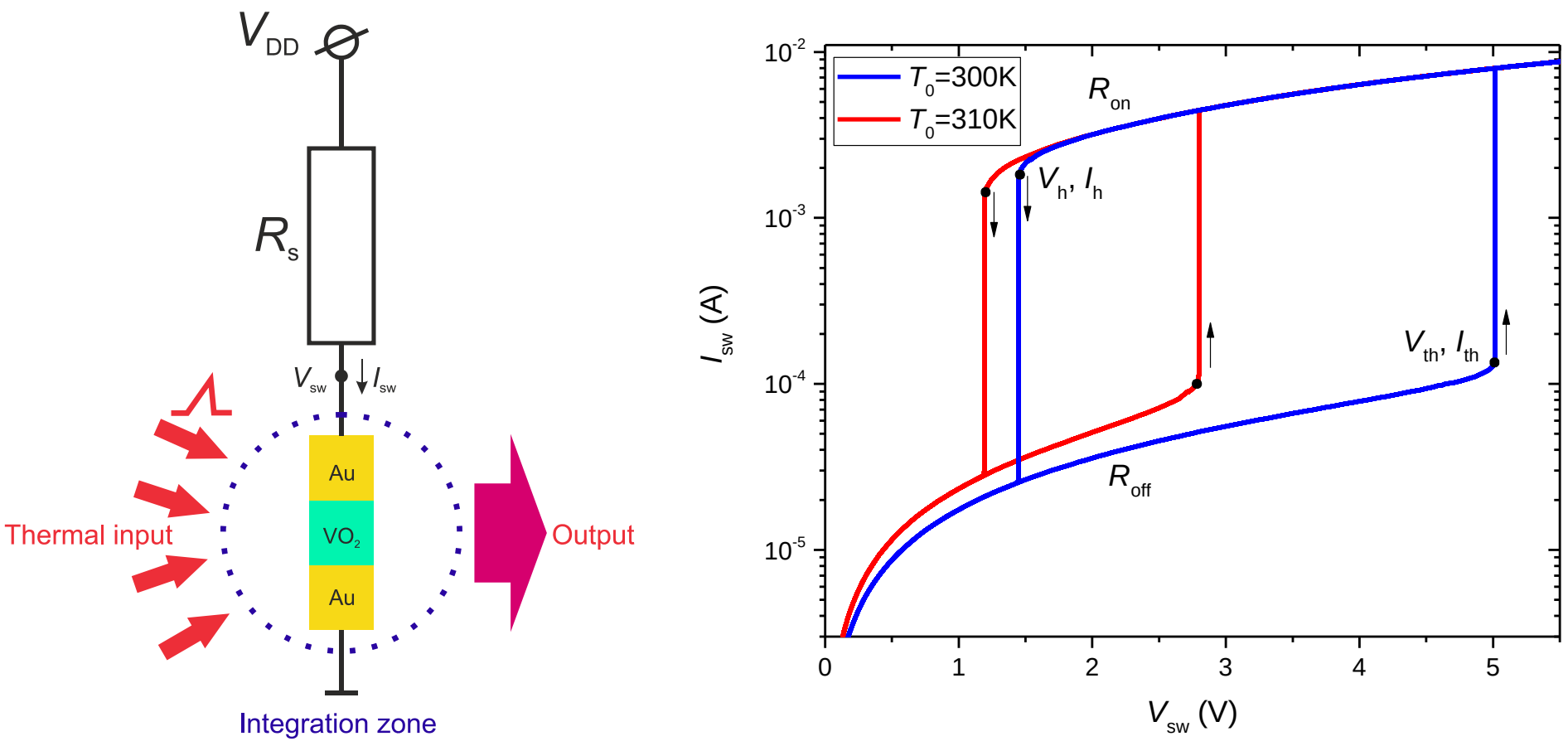

(a) (b)

***Figure 1***. The electric circuit of the $VO_2$ neuron with the thermal mechanism of the membrane potential (a), and the model *I-V* characteristics of the switch obtained at different ambient temperatures $T_0$ = 300 K and $T_0$ = 310 K (b)

Turning the switch on and off is caused by reaching the critical phase transition temperature $T_{th}$ in the $VO_2$ channel region. The temperature of the channel is determined by two competing processes: Joule resistive heating by the flowing current and heat dissipation into the environment. As a result, two branches can be distinguished on the *I-V* characteristic: high-resistance branch ($R_{off}$ ~ 57.6 kΩ), low-resistance branch ($R_{on}$ ~ 630 Ω), switching on point ($V_{th}$ = 5 V, $I_{th}$ = 1.3·10$^{-4}$ A) and switching off point ($V_h$ = 1.45 V, $I_h$ = 1.7·10$^{-3}$ A).

The neuron circuit contains a $V_{DD}$ voltage source, a $VO_2$ switch, and a load resistor $R_s$, connected in series with the switch (Figure 1(a)). In the proposed model, the maximum temperature $VO_2$ of the channel induced by external sources on the region of the switching channel $T_P$ serves as an analogue of the membrane potential. Thermal waves from external sources dynamically change the ambient temperature, increasing it $T_0$' = $T_0$ + $T_P$. The threshold function is the temperature $T_{P_th}$, and the switching effect is observed above this temperature ($T_P \geq T_{P_th}$). The generation by a neuron of a current pulse (spike), which is a one-time transition of a switch from a high-resistance to a low-resistance state, can occur by two mechanisms, presented in Figure 2.

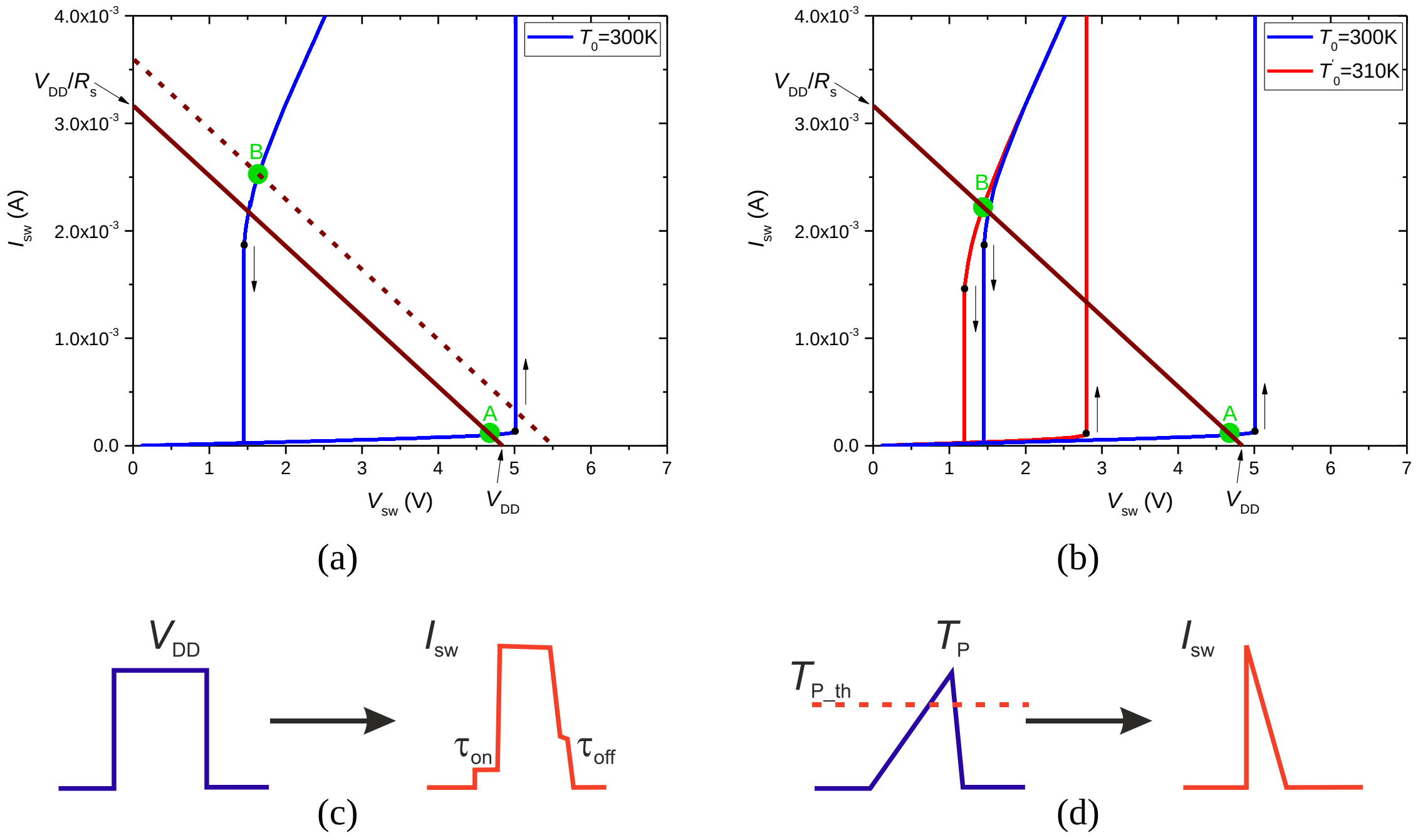

***Figure 2***. I-V characteristic and the load curve during spike generation by the voltage pulse $V_{DD}$ (a, c) and the temperature pulse $T_P$ (b, d)

### *2.1. Mechanism of current spike generation by voltage pulse*

The generation of a current spike by a neuron can happen by the influence of a voltage pulse of supply $V_{DD}$ (Figures 2(a, c)). Figure 2(a) demonstrates the load line of the neuron circuit (Figure 1(a)). It is a straight line of the *I–V* characteristic of the switch, intersecting the ordinate axis ($I_{sw}$) at the point (0, $V_{DD}$ / $R_s$) and intersecting the abscissa axis ($V_{sw}$) at the point ($V_{DD}$, 0). The intersection point of the load line and the *I-V* characteristic is determined by the current flowing in the circuit. This rule is proven by the equation of the neuron circuit:

$$V_{DD} = I_{sw}(V_{sw}) \cdot R_s + V_{sw} \,, \tag{1}$$

where $I_{sw}$ ($V_{sw}$) is the function of the *I–V* characteristic of the neuron. The equation can be solved by the graphical method, presented (1) in the form:

$$I_{sw}(V_{sw}) = \frac{V_{DD}}{R_s} - \frac{V_{sw}}{R_s} \,. \tag{2}$$

The intersection of the *I–V* characteristic $I_{sw}$ ($V_{sw}$) and the load line $I = V_{DD} / R_s - V_{sw} / R_s$ gives a solution to equations (1-2), called the operating point of the circuit.

When the operating point is set to the pre-threshold state (point ***A*** in Figure 2(a)), by the supply of the corresponding voltage $V_{DD}$, the switch is in the closed (high-resistance) state. With a pulsed increase in $V_{DD}$, the load line shifts upward (see dotted notation), a switchover occurs, and the working point moves to a low-resistance branch (point ***B*** in Figure 2(a)). Points ***A*** and ***B*** are stable states of the circuit, as they are located on the branches of the *I–V* characteristic with positive differential resistance. In circuit shown in Figure 1(a), periodic pulsed oscillations can be observed under a constant voltage $V_{DD}$. These oscillations occur if the operating point is on an unstable section of the *I–V* characteristic with a negative differential resistance, between the points ($V_{th}$, $I_{th}$) and ($V_h$, $I_h$).

For a neuron to generate the pulsed signals, a capacitor is usually installed in parallel to the switch [14,15]. However, such capacitors occupy a significant area on the plate; and, to miniaturize the neuron, the capacitors installation should be avoided. Therefore, we used a voltage pulse $V_{DD}$ to generate a current pulse $I_{sw}$. When generating a current pulse $I_{sw}$, the delay times of switching on $\tau_{on}$ and switching off $\tau_{off}$ should be taken into account, which are caused by the time processes of the current channel formation, described in [16]. The delay time $\tau_{on}$ has a strong dependence on the amplitude of the voltage pulse $V_{DD}$, and can vary from a few milliseconds to several nanoseconds.

### *2.2. Mechanism of spike generation by a temperature pulse*

Another mechanism for generating a current spike is the thermal pulse induced on the switch (see Figures 2(b, d)). Temperature waves (pulses) $T_P$, propagating from neighbouring switches, can cause short-term channel heating ($T_0' = T_0 + T_P$), which will lead to a decrease in threshold parameters ($V_{th}$, $I_{th}$, $V_h$, $I_h$). This, in turn, will cause the $VO_2$ switch to switch (Figure 2(b)) from the high-resistance state (point ***A***) to the low-resistance state (point ***B***). Such switching has a threshold character and occurs at $T_P \geq T_{P_th}$. In this case, the value of $T_{P_th}$ can have very small values, fractions of a degree, since point ***A*** can be set arbitrarily close to the switching threshold ($V_{th}$, $I_{th}$), and the dependence of $V_{th}$ on $T_0$ is strong. When heating up by just $T_P$ = 10 K, from a temperature of $T_0$ = 300 K to $T_0'$= 310 K, the value of $V_{th}$ decreases almost twofold. With the decrease of $V_{DD}$, the operating point ***A*** shifts to the left, and the value of $T_{P_th}$ increases. By this way, we can vary the threshold in the $VO_2$ neuron.

## ***3. Modelling the total effect on the membrane potential of the $VO_2$ neuron***

The thermal effect on a neuron was studied using a model circuit consisting of three $VO_2$ neurons, which switching elements "Switch 1, 2, 3" are depicted in Figure 3(a). Neurons 1 and 2 were controlled by a voltage pulse $V_{DD}$ = 6 V and acted on neuron 3, which was supplied with a constant voltage $V_{DD}$ = 1.8 V. The values of the load resistors were as follows: $R_{s_1}$=1 kΩ, $R_{s_2}$= 1 kΩ, $R_{s_3}$=500 Ω. The task was to detect the properties of the integrated action of neurons 1 and 2 on neuron 3. The distances between the switching structures are shown in Figure 3(a). The distance between "Switch 2" and "Switch 3" was larger than between "Switch 1" and "Switch 3" that the effect was not equivalent. We demonstrated in previous studies [17,18] that the temperature wave propagates from the switch along the substrate in all directions at a fixed speed, and the amplitude of the thermal pulse decreases exponentially with distance. By analogy with a biological neuron, a thermal wave corresponds to an action potential propagating along an axon. Using thermal pulses and the dependence of the *I–V* characteristics of the switches on temperature, it is possible to implement a thermal connection between neurons. Moreover, as thermal pulses can overlap each other, forming a kind of interference, a summing input of an LIF neuron can be implemented.

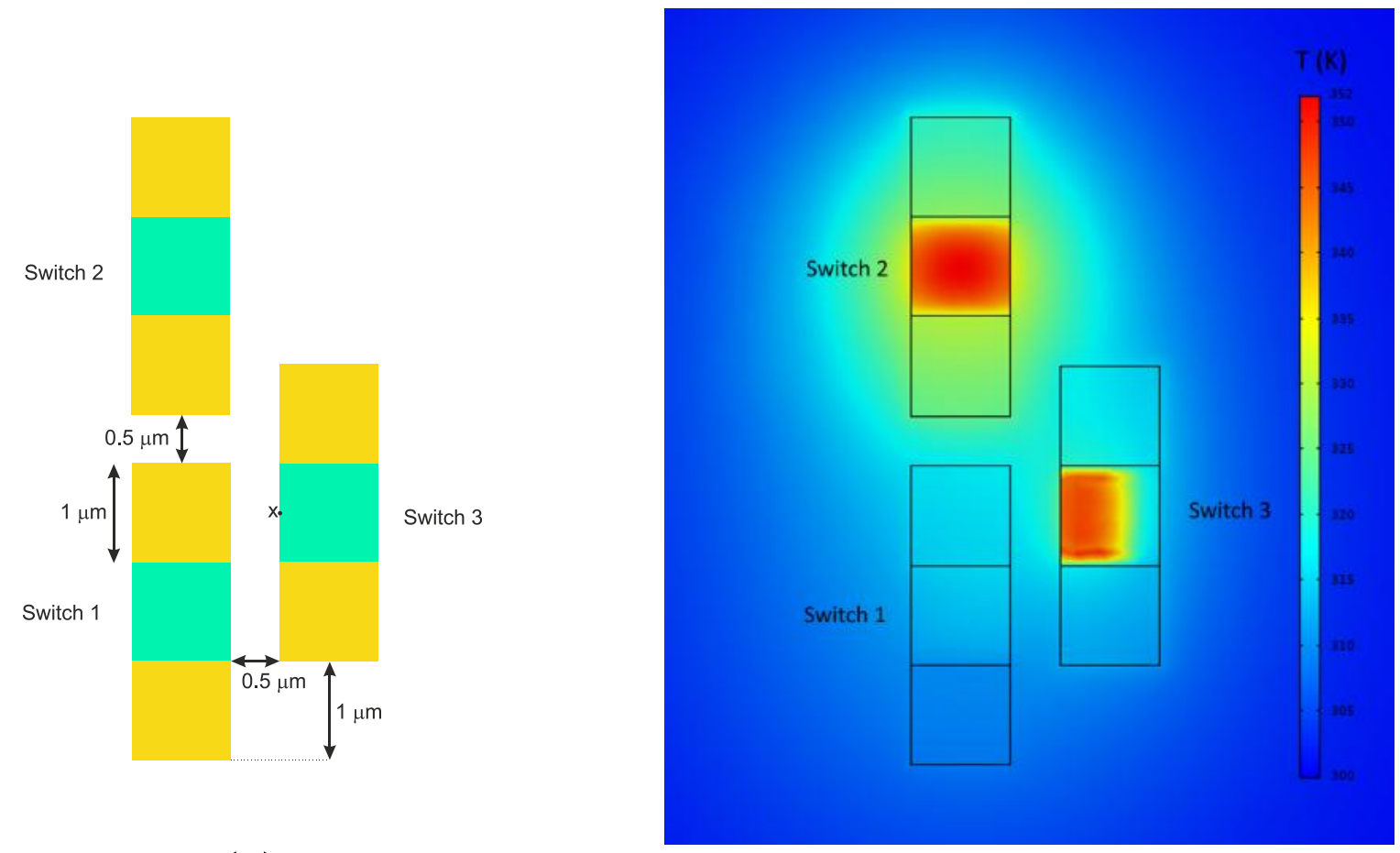

(a) (b)

***Figure 3***. The switches arrangement of the three $VO_2$ neurons (a). The temperature distribution in the modelled structures after the switch 3 is turned on at time $t_1$ in Figure 4(b)

Figure 4(a) demonstrates the thermal effect of neuron 1 on neuron 3. Switching of neuron 1 is initiated by voltage pulse $V_{DD_1}$. According to the current oscillogram, the current $I_{sw_1}$ does not increase and decrease instantly, but there are turn on and turn off times: $\tau_{on} \sim 33$ ns, $\tau_{off} \sim 80$ ns. If the $V_{DD}$ pulse duration is less than $\tau_{on}$, then the switch will not turn on. A similar effect can be used to create a summing effect at the input, and it will be the subject of our future studies. The current in switch 3 ($I_{sw_3}$) increases synchronously with the current $I_{sw_1}$, but it changes slightly, since the induced temperature is insufficient for switching. Figure 4(a) presents the oscillograms of the temperature $T_3$ on switch 3 for two values of $V_{DD_3}$. The measurement point is located on the edge of the channel (point "x" in Figure 3(a)). At $V_{DD_3} = 0$, the dependence $T_3(t)$ characterizes the induced temperature $T_{P_1}(t) = T_3(t) - T_0$ from the first neuron; the graph $T_{P_1}(t)$ is presented in Figure 3(a). The maximum heating temperature is $T_{P_1} \sim 14$ K, and this temperature is not enough for the switching at $V_{DD_3} = 1.8$ V.

With an increase in the duration of the signal supplied to neuron 1, it is possible to achieve switching of neuron 3 (Figure 4(b)). In this case, the induced temperature $T_{P_1}(t)$ exceeds the threshold value $T_{P_th}$ required for switching. Using the dependence of $T_{P_1}(t)$ and the turn-on time, it is possible to determine the value $T_{P_th} \sim 17$ K. The turn-on time is detected by the sharp increase of $I_{sw_3}$ and $T_3$. Therefore, a neuron has a threshold function due to the effect of electrical switching.

If to neuron 1 and neuron 2 the short pulses are fed, which individually cannot initiate the switching of neuron 3, then for certain values of the time shift $\Delta t$ between pulses, we can observe the summation of pulses' temperature effect and the transition of switch 3 to the on state (see Figure 4(c)). The total effect on the neuron $T_P(t)$ can be expressed as the sum $T_P(t) = T_{P_1}(t) + T_{P_2}(t)$, and the graph of the effect is presented in Figure 4(c). Switching occurs, when the threshold $T_{P_th} \sim 17$ K is reached, which corresponds to previous measurements on a single pulse.

The temperature distribution on the switches, at time $t_1 = 1.42$ μs, is presented in Figure 3(b). The maximum temperature and the origin of the conducting channel is observed at the edge of switch 3, from the side of arrival of thermal pulses.

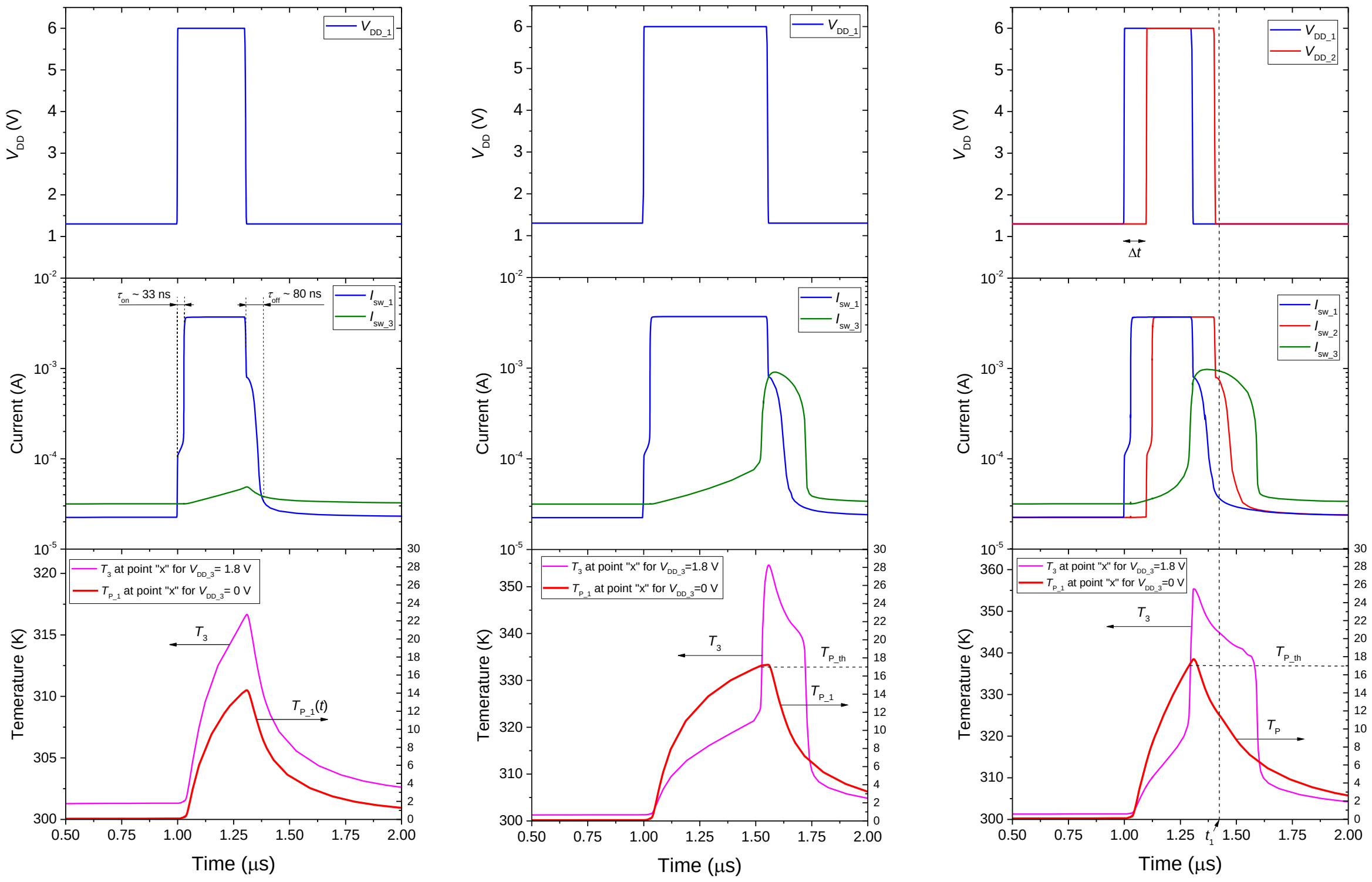

(a) (b) (c)

***Figure 4***. Temporal dependences of voltage, current, and temperature on the switches when applying a pulse of 300 ns (a) and 550 ns (b) duration to switch 1. Supply of pulses with a time shift $\Delta t$ to switches 1 and 2 (c). The time dependences of the corresponding functions of the membrane potential $T_P(t)$ are placed on the temperature scale.

The dependence of the maximum current of switch 3 on the delay $\Delta t$ between two input pulses is presented in Figure 5. This dependence determines the range $\Delta t$ (from -0.3 μs to -0.15 μs), when the pulses "help" each other to turn the switch to the on state. The graph asymmetry with respect to $\Delta t$ = 0 s is caused by different propagation times of thermal pulses from two switches.

In the case of $N$ acting neurons, the membrane potential of the output neuron is calculated as the sum of the temperature effect from all neurons:

$$\begin{aligned} &T_P(t) = \sum_{i=1\ldots N} T_{P_i}(t, r_i) \\ &\text{firing threshold } \ T_P(t) \geq T_{P_th} \end{aligned} \quad . \tag{3}$$

where $r_i$ is the distance to the $i$-th neuron. For short pulses with a duration of ~ 1 μs, the effects associated with the propagation time of the thermal wave are significant and are reflected in phase shift between the thermal pulses induced from the switches at different distances. For example, at a wave velocity of ~ 6 m / s [17], its propagation time by 1 μm is ~ 0.17 μs.

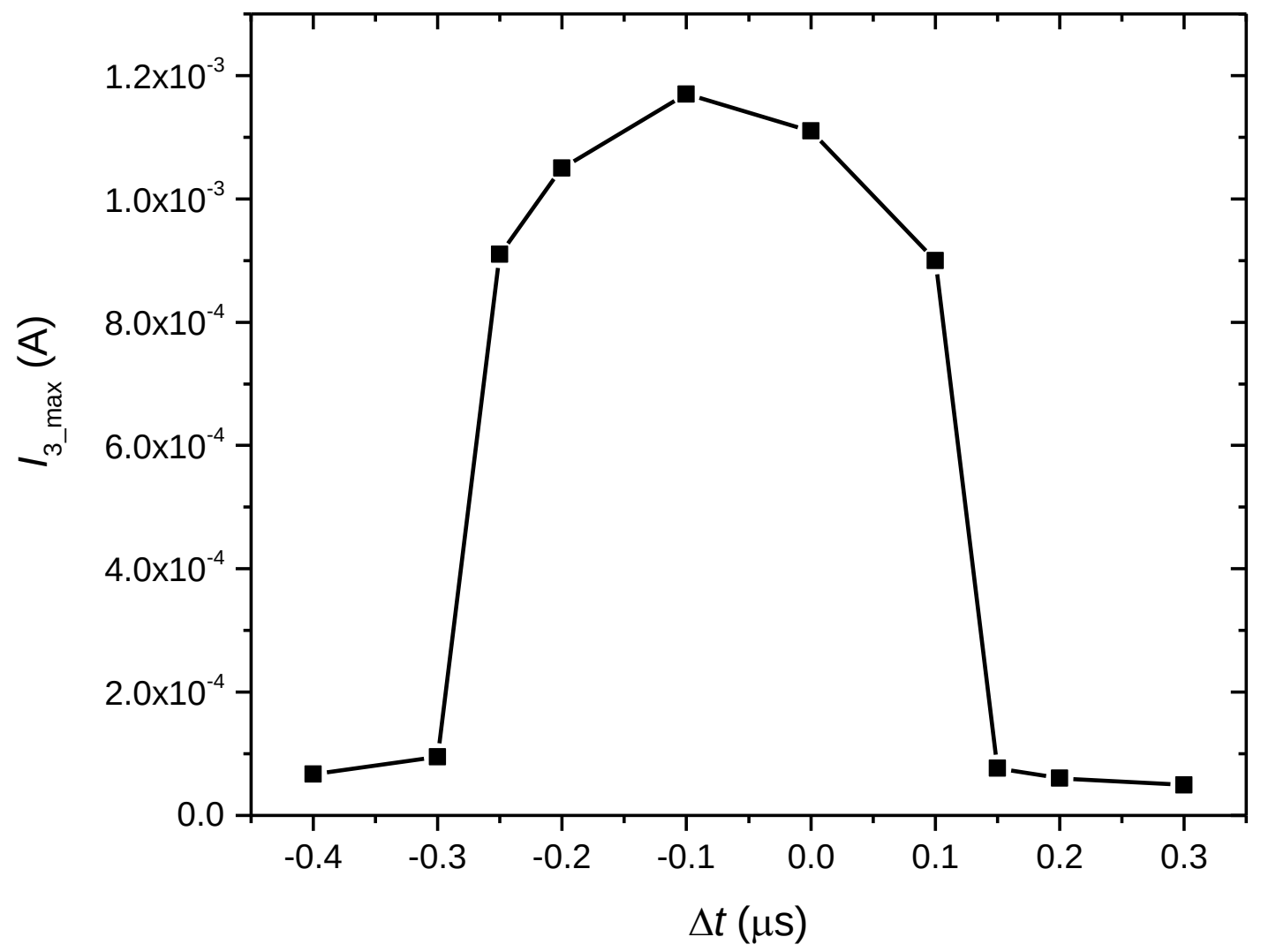

***Figure 5***. The dependence of the maximum current of switch 3 on the delay Δ*t* between voltage pulses applied to switches 1 and 2 (the delay of the second switch relative to the first switch).

## *4. Conclusion*

The study presents a new model of leaky integrate-and-fire $VO_2$ neuron with a thermal mechanism of membrane potential. The operation of three neurons was simulated, and it was demonstrated that the total effect occurs due to interference of thermal waves in the region of the neuron switching channel. The thermal mechanism of the threshold function is caused by the effect of electrical switching, and the magnitude (temperature) of the threshold can vary by external voltage. The neuron circuit does not contain capacitors, making it possible to produce a network with a high density of components, and has the potential for 3D integration due to the thermal mechanism of neurons interaction.

## *Acknowledgment*

This research was supported by the Russian Science Foundation (grant no. 16-19-00135).
The authors express their gratitude to Dr. Andrei Rikkiev for the valuable comments in the course of the article translation and revision.